\documentclass[aps,prl,twocolumn,groupedaddress,showpacs,floats]{revtex4-1}

\usepackage{graphicx,amsmath,amssymb}
\newcommand{\mf}[1]{\boldsymbol{#1}}

\usepackage{color}

\begin{document}

\title{Dipole-Dipole coupled  double Rydberg molecules}

\author{Martin Kiffner${}^{1,2}$}
\author{Hyunwook Park${}^{3}$}
\author{Wenhui Li${}^{1,4}$}
\author{Tom F. Gallagher${}^{3,1}$}

\affiliation{Centre for Quantum Technologies, National University of Singapore, 3 Science Drive 2, Singapore 117543${}^1$}
\affiliation{Clarendon Laboratory, University of Oxford, Parks Road, Oxford OX1 3PU, United Kingdom${}^2$}
\affiliation{Department of Physics, University of Virginia, Charlottesville, Virginia 22904-4714, USA${}^3$}
\affiliation{Department of Physics, National University of Singapore, 117542, Singapore${}^4$}

\date{\today}

\begin{abstract}
We show  that the dipole-dipole interaction between two Rydberg atoms can give rise to  long range molecules. 
The binding potential arises from two states that converge to different separated atom asymptotes. 
These states interact weakly at  large distances, but 
start to repel each other strongly as the van der Waals interaction turns into a resonant dipole-dipole 
interaction with decreasing separation between the atoms. This mechanism leads to the formation of an attractive well 
for one of the potentials. 
If the two separated atom asymptotes come from the small Stark splitting of an atomic Rydberg level, 
which lifts the Zeeman degeneracy, the depth of the well and the location of its minimum are controlled 
by the external electric field. We discuss two different geometries that result in a localized and a donut shaped potential, 
respectively. 
\end{abstract}


\maketitle

A gas of cold Rydberg atoms, atoms in states of high principal quantum number $n$, is more like a solid than 
a room temperature gas. On the 1~$\mu$s time scale of a typical experiment cold, $300\mu$K, Rydberg atoms at a 
density of $10^9$ cm$^{-3}$ only move 2\% of their average interatomic spacing. The atoms are effectively frozen in place. 
Equally important, the atoms interact, due to their large electric dipole transition moments $\mu\approx n^2$, which lead to 
dipole-dipole interactions of magnitude $V_{\text{dd}}=n^4/R^3$, where $\mf{R}$ is the displacement from atom 1 to atom 
2~\cite{tannoudji:api}. We use atomic units unless specified otherwise.  For $n=40$ atoms at $10^9$ cm$^{-3}$, $V_{\text{dd}}=3.8$~MHz, 
and dipole-dipole energy transfer is easily observed in 1~$\mu$s~\cite{anderson, mourachko, noel, ben}. It is closely related to 
the motion of an exciton in a solid and to the Forster energy transfer observed in light harvesting systems~\cite{wuster,engel}. 
The dipole-dipole interactions of Rydberg atoms are also the basis of proposed approaches to constructing quantum 
gates~\cite{jaksch00,lukin}. Their second attraction is that it is possible to form novel long range molecules. 
In what are termed trilobite molecules, a Rydberg atom is bound to a ground state atom by the short range interaction between 
the Rydberg electron and the ground state atom~\cite{greene:00,bendkowsky}. The trilobite molecules are huge, with an internuclear 
separation of $n^2$. In addition, evidence for longer range molecular resonances has been presented, and stable macrodimers, formed by 
van der Waals interactions in weak electric fields, have been observed~\cite{farooqi,overstreet}. The macrodimers 
are huge, with internuclear spacings exceeding $1\,\mu$m. However, they are fragile, existing only in very specific electric 
fields~\cite{overstreet}.

Here we demonstrate that it is possible to form robust long range molecules in which two Rydberg atoms 
are bound to each other by the dipole-dipole interaction. The essence 
of the idea is the following. If there are two sets of dipole-dipole potentials converging to two energetically close 
$R=\infty$ asymptotes, there are potentials converging to the upper asymptote which have wells due to repulsion from 
potentials converging to the lower asymptote. If the two $R=\infty$ asymptotes are produced by lifting the Zeeman 
degeneracy using the Stark shift~\cite{martin3}, the minima of the wells occur at extremely long range, $R>10n^2$. 
In other contexts the Zeeman degeneracy is an unwanted complication~\cite{kiffner:07b,walker}, but here it is an asset. 
A particularly attractive feature of one of these wells is that, for any $n$, the location of its minimum depends on the 
Stark shift. It can, for example, be adjusted to match the separation of ground state atoms in a lattice~\cite{jaksch}, 
offering the prospect of a different way of making an ordered sample of Rydberg atoms~\cite{pohl,raithel}.

After introducing the geometry of the problem we consider the simple case in which the two atoms are aligned in the field 
direction, $\mf{R}\parallel \mf{E}$, for in this case the origin of the well is particularly clear. We then consider the 
more general case in which $\mf{E}$ and $\mf{R}$ are not parallel to show that the well has a finite extent in the angle 
$\theta$ between $\mf{E}$ and $\mf{R}$ and that a well also exists for $\mf{E}\perp \mf{R}$ when the Stark shift is 
reversed in sign. Finally, we comment briefly on the properties, production, and detection of these molecules.

\begin{figure}[t!]
\includegraphics[width=8.5cm]{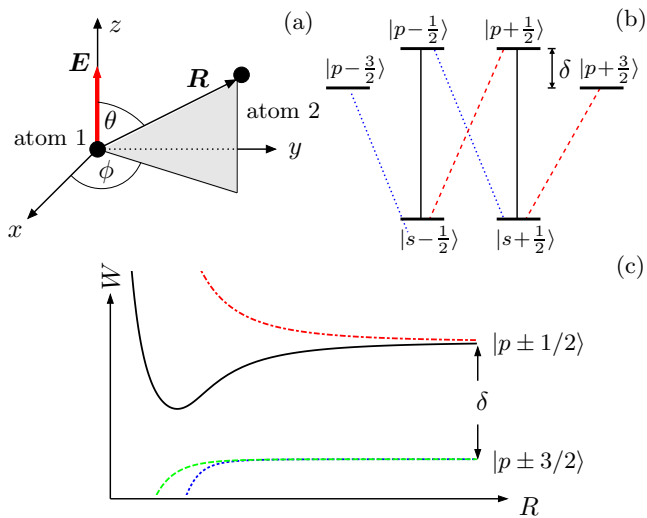}
\caption{\label{fig1}
(Color online) (a) The system under consideration
consists of two Rydberg atoms.
The relative position $\mf{R}$  of atom 2 with respect to atom 1
is expressed in terms of spherical coordinates. An external electric field $\mf{E}$ is applied
in the  $z$ direction.
(b) Internal level structure of each Rydberg atom.   The Stark shift $\delta\equiv W_{p \pm1/2}-W_{p \pm3/2}$ 
is positive for the above configuration.
Only states connected by solid, blue dotted and red dashed lines are dipole coupled. 
(c) Potential curves for the $M=1$ $nsnp$ states as a function of the internuclear spacing $R$. 
The Stark energy difference $\delta$ between the $|p\pm1/2\rangle$ and $|p\pm3/2\rangle$ states is assumed to be positive.
}
\end{figure}
The geometry of the two atom system under consideration is shown in Fig.~\ref{fig1}a. 
The electric field $\mf{E}$ is in the $z$ direction, 
which we choose to be the axis of quantization, and $\mf{R}$ is at the angle $\theta$ from the $z$ axis. We here consider 
$ns_{1/2}$ and $np_{3/2}$ states, but the same reasoning applies to other pairs of dipole-dipole coupled states. It is 
convenient to specify the atomic states by their orbital angular momentum $\ell$ and azimuthal total angular momentum $m_j$, 
i. e. as $|\ell m_j\rangle$, for example, $|p +3/2\rangle$. The primary effect of the electric field is to lift the Zeeman 
degeneracy of the $p_{3/2}$ state so that $W_{p \pm1/2}\neq W_{p \pm3/2}$, although $W_{p m_j}=W_{p -m_j}$. As a result, 
there are two sets of dipole-dipole coupled states with nearly degenerate $R=\infty$ asymptotes, which leads to the potential 
wells. The small spin-orbit splittings of the higher $\ell$ states should also lead to wells, but their positions will be fixed. 
We define the energy splitting $\delta$ by $\delta\equiv W_{p \pm1/2}-W_{p \pm3/2}$. By the appropriate choice of AC or DC 
electric field $\delta$ can be made positive or negative~\cite{martin3}.  In Fig.~\ref{fig1}c we show the $\delta>0$ case in which 
the $|p\pm1/2\rangle$ states lie above the $|p\pm3/2\rangle$ states at $R=\infty$.

The system of interest is the diatomic system of one $ns_{1/2}$ atom and one $np_{3/2}$ atom, and we term its states the 
$nsnp$ states. We describe the $nsnp$ states as ordered direct products of atomic states. For example, atom 1 in the 
$|s+1/2\rangle$ state and atom 2 in the $|p+3/2\rangle$ state yields the state $|s +1/2,p+3/2\rangle$. Constructing all the 
possible direct products of $ns_{1/2}$ and $np_{3/2}$ states provides a complete set of basis functions for the $nsnp$ states. 
With our choice of quantization axis the total azimuthal angular momentum $M$ of the system remains a good quantum number 
in the presence of the electric field. There are four states of $M=0$, eight of $M=\pm1$, and four of $M=\pm2$. 
We first consider the case in which the two atoms lie on the $z$ axis, that is $\mf{R}\|\mf{E}$, and $\delta>0$. 
The $nsnp$ states have two asymptotic $R=\infty$ energies separated by $\delta$, as shown by Fig.~\ref{fig1}c. 
When $\mf{R}||\mf{E}$ $M$ remains a good quantum number in the presence of the dipole-dipole interaction, 
which is given by~\cite{tannoudji:api}
\begin{align}
\hat{V}_{\text{dd}}= \frac{1}{ R^3}[\mf{\hat{d}}^{(1)}\cdot\mf{\hat{d}}^{(2)}
-3(\mf{\hat{d}}^{(1)}\cdot\vec{\mf{R}})(\mf{\hat{d}}^{(2)}\cdot\vec{\mf{R}})], 
\label{vdd}
\end{align}
where $\mf{\hat{d}}^{(i)}$ is the electric dipole-moment operator of atom $i$ and $\vec{\mf{R}}=\mf{R}/R$ 
is the unit vector along the molecular axis. 
For any finite $R$ the dipole-dipole interaction only couples states of the 
same $M$, and we focus on the four $M=1$ states. We ignore the $M=0$ and $\pm2$ states. There are four $M=-1$ states degenerate with 
the $M=1$ states, and the reasoning for the $M=1$ states also applies to them. As shown by Fig.~\ref{fig1}c, as $R\rightarrow\infty$ 
there are two $M=1$ states which converge to the $|p\pm1/2\rangle$ asymptote and two converging to the  $|p\pm3/2\rangle$ asymptote. 
As $R\rightarrow\infty$ the former two states have dipole-dipole energy shifts proportional to $\pm1/R^3$, while the latter pair 
are not dipole-dipole coupled to each other and have no first order dipole-dipole interaction. At small $R$, where 
$V_{\text{dd}}\gg|\delta|$ two of the four $M=1$ levels are shifted up in energy and two down. Since the $M=1$ levels can not cross, 
the lower level connected to the  $|p+1/2\rangle$ asymptote must have a well, as shown in Fig.~\ref{fig1}c, and the minimum occurs at a 
value of $R$ such that $V_{\text{dd}}\approx|\delta|$. 
For the evaluation of Eq.~(\ref{vdd}) we compute the matrix elements of the dipole operator 
via the  the Wigner-Eckart theorem~\cite{edmonds} according to 
\begin{align}
 \langle p m |\mf{\hat{d}}^{(i)}|s m' \rangle = 
\underbrace{\frac{(np_{3/2}\|\mf{\hat{d}}\|ns_{1/2})}{[2\cdot3/2 +1]^{1/2}}}_{=\mathcal{D}} 
\ \sum_{q=-1}^1 C^{3/2 m}_{1/2 m' 1 q} \vec{\mf{\epsilon}}_q, 
\label{mele}
\end{align}
where  $C^{3/2 m}_{1/2 m' 1 q}$ are Clebsch-Gordan coefficients 
and $\vec{\mf{\epsilon}}_q$ are orthonormal unit vectors 
arising from the decomposition of the dipole operator into its spherical components~\cite{edmonds}. 
The reduced matrix element  $(np_{3/2}\|\mf{\hat{d}}\|ns_{1/2})$ in Eq.~(\ref{mele}) 
can be written in terms of a radial matrix element between the  $np_{3/2}$ and $ns_{1/2}$ states~\cite{edmonds,park,walker}. 
We find $\mathcal{D}=\sqrt{1/3}\langle np|r|ns\rangle$, and 
for alkali atoms $\langle np|r|ns\rangle \cong n^2$ for $n\sim40$~\cite{walker}. 
Since the sum in Eq.~(\ref{mele}) is a term of order unity, the magnitude of $V_{\text{dd}}$  is given by 
\begin{align}
 \Omega=|\mathcal{D}|^2/R^3. 
\end{align}
Equating $\Omega$ to $|\delta|$ yields the characteristic length $R_0$, given by
\begin{align}
R_0=(|\mathcal{D}|^2/|\delta|)^{1/3}=[n^4 /(3 |\delta|)]^{1/3}.
\label{r0}
\end{align}
Fig.~\ref{fig1}c also suggests that the depth of the potential well is approximately $|\delta|/2$.

\begin{figure}[t!]
\includegraphics[width=8.6cm]{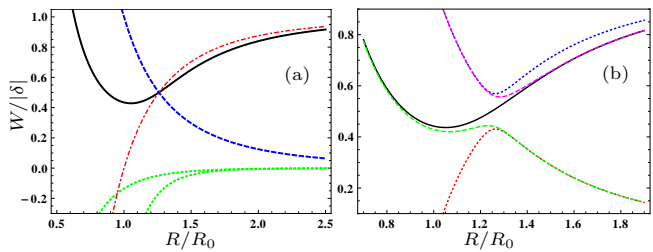}
\caption{\label{fig2}
(Color online) (a)The $\theta=0$ $M=\pm1$ potentials and the $M=0,\pm2$ potentials which cross them. 
The potential well (black solid line) is two-fold degenerate. The blue dashed
line corresponds to the $\text{M}=\pm 2$ states, and this energy curve is two-fold degenerate.
The red dashed-dotted line corresponds to a single  M=0 state. The dotted green lines correspond to the van der Waals shifted
$\text{M}=\pm 1$ states, and each of them is two-fold degenerate. 
(b) Expanded view of the point of five-fold degeneracy in Fig.~\ref{fig2}a for
a small angle $\theta=5^\circ$. Note that the figure looks qualitatively the same
for even smaller angles. There is now a single state corresponding to the energy well
(black solid line).
}
\end{figure}

\begin{figure}[b!]
\includegraphics[width=7.5cm]{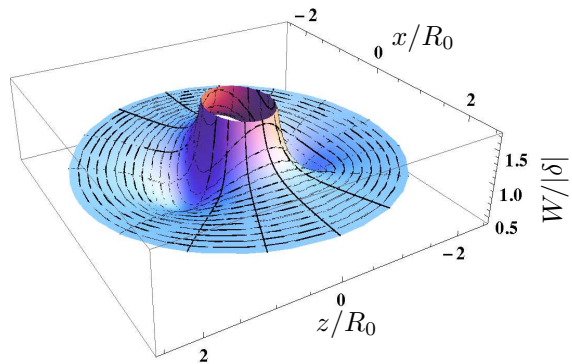}
\caption{\label{fig3}
(Color online) Potential of the $\theta=0$ well state in the $x$-$z$ plane for $\delta>0$.
Two pronounced minima occur on the $\pm z$ axes. The potential is
azimuthally symmetric around the $z$ axis.
}
\end{figure}

While it is clear that there is a potential well when $\theta=0$, two important questions remain. First, when $\theta\neq0$ 
the dipole-dipole interaction couples states of different $M$, and we can no longer ignore states of $M\neq1$ which cross 
the $M=1$ states. What is the effect of these couplings? Second, what is the angular extent of the well in $\theta$? Is it 
large enough to be useful? To address these questions we calculate the energy levels as functions of $R$ and $\theta$ by 
diagonalizing the Hamiltonian matrix resulting from the Stark shift $\delta$ and the dipole-dipole interaction $\hat{V}_{\text{dd}}$. 
We first consider the effect of the other $M$ states. The dipole-dipole matrix elements have the following dependence on $\theta$.
$\Delta M=0$: $1-3\cos^2\theta$, $\Delta M=\pm1$: $\sin\theta\cos\theta$, $\Delta M=\pm2$: $\sin^2\theta$. 
In general, all the levels are coupled. In Fig.~\ref{fig2}a we show the calculated $M=\pm1$ levels for $\theta=0$ as well as the $M=0$ 
and $\pm2$ levels which cross the well level. For clarity we have omitted the other $M=0$ and $\pm2$ levels. 
In Fig.~\ref{fig2}b we show 
the energy levels for $\theta=5^\circ$. All the levels are coupled, and the degeneracies of the $M=\pm1$ and $M=\pm2$ states 
are broken. Equally clear, there are several potential wells, one of which has the same shape as the one shown in Figs.~\ref{fig1}c 
and~\ref{fig2}a. Calculating the energy levels as a function of $\theta$ enables us to determine the angular extent of the well shown 
in Fig~\ref{fig2}b, and in Fig.~\ref{fig3} we show a plot of the potential vs the $x$ and $z$ directions. 
The potential is azimuthally symmetric, 
and the potential wells exist along the $+z$ and $-z$ axes with widths of approximately $20^\circ$, wide enough to be useful.

\begin{figure}[t!]
\includegraphics[width=7.5cm]{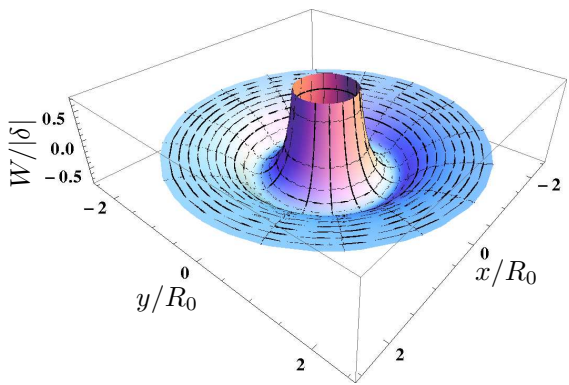}
\caption{\label{fig4}
(Color online) Potential of the $\theta=\pi/2$ well state in the $x$-$y$ plane for $\delta<0$.
The angular width of the well state around  $\theta=\pi/2$ is similar to
the width of the corresponding well state for $\delta>0$ around $\theta=0$, see Fig.~\ref{fig3}.
}
\end{figure}

If we reverse the Stark shift, so that $\delta<0$ and the $|p+3/2\rangle$ level lies above the $|p+1/2\rangle$ level, and 
consider $\theta=\pi/2$, there is also a well. For $\theta=\pi/2$, $M=\pm1$ states are coupled to each other but not to 
the $M=0$ and $\pm2$ states. We again focus on the $M=\pm1$ states, ignoring the $M=0$ and $\pm2$ states. There are $M=\pm1$ 
states converging to both $R\rightarrow\infty$ asymptotes, and the well resembles the well of Figs.~\ref{fig1}c and~\ref{fig2}a, 
but it is nondegenerate in this case. Unlike the $\theta=0$ case, the strong $\Delta M=\pm2$ coupling lifts the degeneracy of 
the $M=1$ and $M=-1$ states. Due to the azimuthal symmetry the well is donut shaped as shown in Fig.~\ref{fig4}. Calculating 
the energy levels as a function of $\theta$ shows that the angular width of the potential about $\theta=\pi/2$ 
is $\pm20^{\circ}$, similar to the well shown in Fig.~\ref{fig3}. 

Analytic diagonalization of the Hamiltonian matrix yields an expression for the potential wells shown in Figs.~\ref{fig2}b 
and~\ref{fig3}. Explicitly,
\begin{align}
\label{well}
 V(R)= & \frac{1}{6}\big[3\delta -4 \Omega   \\
& +\sqrt{28 \Omega^2 +3 \delta ( \Omega + 3 \delta - 9 \Omega  \cos(2\theta)})\big]. \notag
\end{align}
This formula is valid for $\delta>0$ and $\delta<0$. From the potential of Eq.~(\ref{well}) it is straightforward to derive 
expressions for the position of the minimum of the potential and the vibrational frequency of the radial motion in 
the potential. For $\theta=0$ the minimum of the potential occurs at
\begin{align}
 R_{\text{min}}=(7/6)^{1/3} R_0, 
\label{rmin1}
\end{align}
and the well depth is $\Delta V=0.57|\delta|$.
The frequency of the vibrational motion is approximately given by
\begin{align}
 \omega_{\text{vib}}= 2 \sqrt{\frac{|\delta|}{R_0^2 \mu}},
\label{rvib1}
\end{align}
where $\mu$ is the reduced mass of the two atoms.

For $\delta<0$ and $\theta=\pi/2$ the minimum of the potential occurs at
\begin{align}
 R_{\text{min}}=(4/3)^{1/3} R_0,
\end{align}
and the potential depth is $\Delta V=0.75|\delta|$.
The frequency of the vibrational motion is approximately given by
\begin{align}
 \omega_{\text{vib}}= 2.4 \sqrt{\frac{|\delta|}{R_0^2 \mu}}.
\end{align}
The Stark splitting of the $np_{3/2}$ level must be small compared to the $np$ fine structure interval, which in 
Rb and Cs is $\sim0.01n^{-3}$~\cite{gallagher}. Accordingly, a reasonable limit for $\delta$ is $|\delta|<10^{-4}n^{-3}$. 
Combining this limit with Eq.~(\ref{r0}) yields
$R_0>(10^{4}n/3)^{1/3} n^2$,  
which is substantially larger than the Rydberg atom.

To develop a feeling for the dynamics in these shallow potential wells, we use Rb $n=40$ atoms with 
$\delta/2\pi=10$~MHz as an example. Combining these values with Eqs.~(\ref{r0}),~(\ref{rmin1}) and~(\ref{rvib1}) leads to 
$R_0=8.25\times10^4$, $R_{\text{min}}=8.69\times10^4$, and $\omega_{\text{vib}}/2\pi=24$kHz, 
so motion on the potential curves is adiabatic. Since an equal linear superposition of the Rb $40s$ and $40p$ states 
has a 0 K radiative lifetime of 120~$\mu$s, only three vibrational oscillations are likely~\cite{gallagher}. 

The two atom states in these potential wells can be produced by laser excitation of pairs of atoms to the diatomic $nsns$ 
state with subsequent excitation to the dipole-dipole bound $nsnp$ state by microwave excitation. While the most interesting 
prospect is forming the dipole-dipole wells to match the spacing of atoms in a lattice~\cite{jaksch}, it is useful to consider 
other ways in which these wells can be manifested.  Definitive evidence can be obtained by taking advantage of the fact that in 
zero field the potentials are either attractive or repulsive. There are no long range wells in which atoms can be trapped, and 
atoms are not stable at energies slightly removed from the $R=\infty$ energies. Stable atoms in these potential wells can be 
detected in several ways. First, driving a second microwave transition from the $nsnp$ well to, for example, the $nsn'd$ 
state, which has no dipole-dipole interaction, should show a satellite feature removed from the atomic transition by the well 
depth. This feature should become more pronounced if a time delay is introduced before the second microwave transition, the 
limiting factor being radiative decay. Second, the motion of atoms on attractive dipole-dipole potentials in zero field leads 
to collisional ionization on time scales much faster than the radiative lifetime~\cite{li,amthor, viteau, park2}. This 
ionization should be significantly suppressed by the existence of the long range potential dipole-dipole wells. for example, 
in the $\theta=\pi/2$ case with $\delta<0$, which leads to the donut potential of Fig.~\ref{fig4}, only one of the four potentials at 
energies just below the upper $R=\infty$ asymptote is purely attractive and can lead to collisional ionization; the other three 
have potential wells. Finally, it may be possible to detect the broadened time of flight field ionization signals 
due to atoms which remain close together~\cite{overstreet}.

In conclusion, the Stark tuned dipole-dipole potentials described here provide a novel, adjustable way to locate interacting 
atoms at a chosen distance.

\begin{acknowledgments}
It is a pleasure to acknowledge useful discussions with D. Jaksch and R. R. Jones, the support of the Air Force Office of 
Scientific Research and of the National Research Foundation and the Ministry of Education, Singapore. 
\end{acknowledgments}

\end{document}